# The Next Frontier for Openness: Wireless Communications

**Early Discussion Draft**

*Eli M. Noam*

*Professor of Finance and Economics, and Director, Columbia Institute for Tele-Information, Columbia University*

*Paper for the 2001 Telecommunications Policy Research Conference*

*Alexandria, Virginia*

September 25, 2001

**Abstract**

For wireless communications, the FCC has fostered competition rather than openness. This has permitted the emergence of vertically integrated end-to-end providers, creating problems of reduced hardware innovation, software applications, user choice, and content access. To deal with these emerging issues and create multi-level forms of competition, one policy is likely to suffice: a *Carterfone* for wireless, coupled with more unlicensed spectrum.

**The Wireless Policy Mess**

Openness is more than competition. Competition means the ability of companies to contest each other and to seek customers' business. This can result in efficiency and enhanced consumer welfare. But it can also result in a competition among bundled product packages instead of competition on a product-by-product basis. Openness, on the other hand, means the ability of competitors to access consumers directly rather through their own rivals. This is particularly an issue in network industries, and has been a constant theme of regulatory battles for over a century. In telecommunications, product and service markets were closed to competitors for a long time. For example, rival equipment makers existed domestically and internationally but could not reach customers of AT&T" network. Telephone networks were opened first to customer equipment. Then it was long distance and international service. Then, network equipment. Then, local telecommunications. It has now been extended to internet service over cable TV networks. But it has not yet reached wireless communications. Here, competition has been fostered, but not openness. To the contrary, most trends of wireless policy are in the opposite direction. But this is not well understood.

In consequence, this article takes issue with the basic philosophy of the FCC's wireless policy. What makes this policy disturbing is: (a) that it does not seem to benefit from the positive impulses of other policy reforms; (b) that it cannot be explained as based on political constraints, as in the case of broadcast TV; and (c) that it largely focuses on carriers' flexibility rather than users' choice menu.

It might be objected at the outset that wireless has been proceeding in the US by leaps and bounds, so why criticize an evident success story? To be successful in wireless is not very difficult.  Wireless is a success story everywhere in the world.  Growth rates in Cambodia, Bulgaria, and Zimbabwe are phenomenal, too, and not necessarily due to enlightened government regulation, but rather due to a universal demand for ubiquitous communications and a supply of advanced signal processor chip technology.  Governments can hardly claim credit for these developments, though that has not stopped them, of course.

In the US, if anything, government has been the brake, not the engine.  First, the original duopoly system for the analog first generation took many years to establish, and during that time the US lost its original technology head start to the Scandinavians, who still benefit from those early years.  It has been estimated that the delay cost the US $20 billion.  This is an underestimate, since it counts only lost carrier revenues, not lost consumer welfare, productivity gains, exports, and jobs. In Finland, meanwhile, Nokia accounts for one quarter of all national export earnings.

When it came to the second generation of mobile communications, US policy was again painfully slow and complex.  Some of this cannot be helped, being due to a lawyer-saturated and player-congested environment, but partly it was self-inflicted.  The reason was the creation of a system of spectrum allocation of mind-boggling complexity, whose aftermath is still winding its way to the Supreme Court, almost a decade later.  This system centered on spectrum auctions with upfront payments, a system much beloved to game-theorists without much knowledge of the wireless environment, by property rights ideologues, and by government officials eager to fill the empty coffers of government with windfall revenues. They sold long term assets for current consumption.  They sincerely believed that they had created an efficient form of allocation when in reality they established an extraordinary form of entry barriers to new entrants, and an upfront entry tax on established companies which were given the choice, in effect, to pay or die. I have discussed this in greater detail elsewhere.[1]

The creation of this system generated costs of delay far in excess of any efficiency gains. It permitted most of Europe and Japan to forge further ahead in technology, applications, and consumer satisfaction. Many millions of i-mode or short messaging users roam the streets of Japan and of Europe, but not of the US. This is not to claim that the US is a backward country when it comes to wireless. It is not. But one must note that in almost all other fields of communications the US is heavily dominant. Why not in mobile wireless? This is, after all, a which should be a natural field of leadership, given the American love affair with mobility and dependence on the automobile. The one different variable is policy. The one pleasant exception is the emergence of a standard –CDMA—that has been a real advance. But even this case benefited more from US defense technology than communications policy.

In the now emerging third generation of wireless, US policy is again slow and uncertain. Allocation of spectrum has been a near-farcical process of bargaining among entrenched industries and bureaucracies. Fortunately, the Europeans and Japanese have encountered problems of their own which permit us to pretend that we have engaged in a process of grave policy deliberation, instead of simply being unable to get our house in order. It should also be noted that one of the main problems Europeans have encountered is due to the auction with upfront payment

---

[1] See Noam, Eli M., "Spectrum Auctions: Yesterday's Heresy, Today's Orthodoxy, Tomorrow's Anachronism. Taking the Next Step to Open Spectrum Access," *The Journal of Law & Economics*, vol. XLI part 2 pp. 765-790 (October 1998).

process, a successful US export that had received the eager attention of European budget officials.

Each of these setbacks can be explained. Collectively, they raise the question if we are proceeding with the right strategy, or whether we have the fundamentally wrong approach. It is rare to find European telecommunications policy being more pro-openness and pro-consumer choice than American one[2], but this is the case for wireless communications.

American telecommunications and information policy has been at its strongest when it focused on consumer choice and on the lowering of entry barriers. This translated to a willingness to let control over communications shift from the core of the network to the periphery, and for the core of the network to be competitive. The internet is the classic manifestation of this philosophy. Its success, in contrast to government-sponsored, centralized, PTT-driven videotex operations such as the Minitel, btx, Captain, Prestel, etc, has demonstrated the fundamental strength of this model.

It is therefore regrettable that it appears that the FCC has not applied the lessons from past successes to wireless, too. But it is never too late. A new crew is at the oars and tiller, and it might take a new look before it becomes responsible for another $20 billion plus in foregone future benefits.

**The Problem of Vertical Integration**

Mobile communications are becoming the front-line communications device for most people. Ubiquitous, always on. In the recent attacks on American cities, it was used from airplanes, from under the rubble, as a substitute for congested landlines. Already, over 120 million Americans subscribe to cellphone service, almost as many as to wireline telephony. It will soon overtake the latter, as it already has in several countries. Wireless is moving into internet access, transactions, and media content. It is becoming too important to ignore.

The major problem with the emerging wireless environment is that it is vertically integrated in ways that are now unthinkable in other media. Could one imagine a telephone carrier that can limit user access only to its own internet portal, that can select the websites accessible, that can control the type of telephone equipment its users are attaching, and the software these users are downloading? This has not been particularly clear in the past, where cell phones could be thought of as some kind of advanced cordless phone for the car. But they are now becoming much more than that, and for more people.

Furthermore, American cellular communications are ceasing to be a wide-open field. Carriers have consolidated to 6 national footprints – AT&T, Cingular, Nextel (in regulatory terms not quite a cellular provider), Sprint, Verizon, and VoiceStream. These companies use, between themselves, 5 different standards, one of them the analog AMPS that is on its way out; and four digital standards -- TDMA, CDMA, GSM, and IDEN. These carriers are widely expected to further consolidate into four firms. They are likely to continue using several different standards, with only limited transferability of handset equipment among carriers. Customers are therefore limited in the ability to easily switch carriers. These different standards create *consumer lock-in*, and permit each of the carriers to maintain a stronger market power over its customers. Competition in technology has, ironically, reduced carrier competition for consumers.

---

[2] See Eli M. Noam, *Telecommunications in Europe,* Oxford University Press, 1988.

The main characteristic of the wireless business, which affords the carriers such lock-in over its customers, is vertical integration. The customer is a contractual subscriber who is served vertically by a wireless carrier that provides a full set of services. The basic components of a wireless operation are graphed in the following:

**Graph 1.**

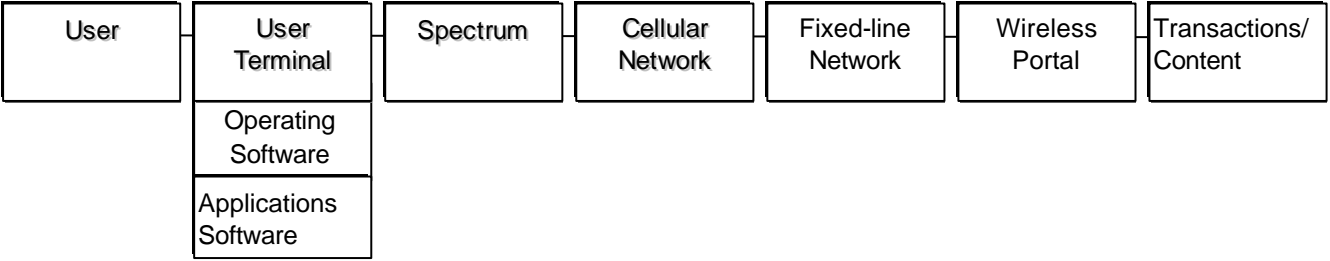

The user reaches his carrier via his terminal (handset) and the carrier's allocated frequencies. The call is then routed via a fixed network to the recipient. More recently, this has been extended to route internet-style communications to a wireless portal that links into transactions and content.

The key component to this system is the spectrum allocation. It enables the carrier to control downstream the terminal equipment and access of a subscriber, and leverage this position of "owning the customer" upstream to the other steps of this chain.

In consequence, we are quite used to the notion that the carrier:

- Selects, markets, and approves the customer handsets and connects it to its network
- Provides, selects, and adopts many of the features, capabilities, and content resident on the handset
- Operates the wireless portion of the communications path
- Operates or provides the local fixed line distribution
- Operates or selects the long distance and international carrier
- Selects, for areas in which it does not provide service itself, a partner mobile carrier that services the subscriber, at rates negotiated and billed by itself
- Provides software-defined functionalities on the network
- Selects and approves services resident on the network and provided by itself or by third parties
- Controls the access to a wireless portal, its content and features, of the providers linked by that portal, and of the placement of these links

There is nothing wrong with a carrier offering all of these components in a bundled fashion. However, when there is no alternative to taking the entire bundle or major parts of it, several problems are inevitable. One can readily recognize good old friends, issues that have bedeviled fixed line telephony and cable television:

- The reduction or lack of customer choice in applications and content inherent in a vertical integration with no or limited alternatives
- The reduction in innovation of service provision due to the closed nature of the applications and software that can be offered by third parties

- The absence of choice for customers to use, where more advantageous, alternative wireless arrangements are possible, such as wireless LANs, other carriers for roaming, or stronger signals of another carrier
- Market power with respect to vendors of m-commerce, and requirements on such vendors to become business partners
- Selectivity over content, which would be particularly troubling if the wireless medium would become a mass medium with video, audio, and text
- Restrictiveness in the inter-carrier transfer of instant messaging.

These problems will now be analyzed in greater detail.

**The Problems of Non-Openness**

*1. Reduction of Choice among cellular service providers*

Right now, cell phone users enter into a service agreement with a single carrier. That carrier accepts all of their calls, or reaches them in the case of incoming calls. Where the user is outside the service territory of the carrier, the user gets serviced by another carrier in a "roaming" arrangement. The roaming-partner carrier is selected by the primary carrier in a commercial agreement (a "preferred" roaming arrangement). The call could also be picked up whichever carrier is around (a "general" roaming partner, typically a set of carriers, with prices set industry-wide). The third type of arrangement is based on signal strength, where the roaming goes to the strongest signal in that area, unless there is a primary or general roaming agreement, which would override. Whatever the arrangement, the user has no choice in the matter, in contrast to the arrangement in GSM countries, where a caller can select the roaming carrier and override its primary carrier's choice.

In the US, this choice is further limited by the different wireless protocols used by carriers. A user of a carrier operating on the CDMA standard cannot roam, in technical terms, on a TDMA or GSM carrier. A limited number of handsets can use both TDMA and GSM since they are related. But on the whole, the ability to switch to a carrier using another standard is minimal. In contrast, in GSM countries users can easily take their handset to any other carrier.

Furthermore, it is impossible to subscribe to more than one carrier using a single handset. For example, if a user spent much of his time in both New York and Atlanta, and no company serviced both cities, he might want to subscribe to companies in both cities rather than pay expensive roaming charges. However, there is presently no practical possibility to switch between two carriers. In theory, something exists called "dual NAM" that would permit dual-carrier subscriptions. In practice, however, phone inquiries to several major carriers did not reveal the availability of such arrangements.

Also in theory, a reseller or reseller group could resell the services of more than one carrier or service type. This assumes that permission would be granted by the carriers whose service is being resold, which is not likely if they refused to permit such choice for their direct customers.

This lack of choice has real implications. Roaming calls are quite expensive, and are not part of the subscriber's "bucket" of minutes. They are a major moneymaker for carriers.

The main problem here is not technology but resistance to competition. Once a user can switch freely among carriers, where will it end? A user might regularly drive through some areas where the signal of his primary carrier is missing, and then select another carrier that performs better. Next, a user might switch to a carrier who offers her the lowest rate during that time period. Soon, the user would be able to engage in "least cost routing". This means that there might be automated competition for every call, as opposed to the present system of competition for the subscription.

*2. Absence of Choice Among different Wireless Services*

In the past, cellular phone service constituted an end-to-end service, separate from those of others. However, other wireless services are also being offered. Paging has long been a widespread service, and smart paging via narrowband PCS has gained increasing popularity. An example is the BlackBerry pager for always-on email. Some such services are being offered on cellphone terminals, but only using its cellphone frequencies, as opposed to being able to switch to the service provided by another paging company. Furthermore, a cell phone terminal could conceivably be used as a terminal for a cordless phone at home or at the office, directly without going through the wireless network. Similarly, it could be used as a "walkie-talkie" between several other cell phones in a neighborhood, again without going through the actual network. (This is a popular feature provided by Nextel for its own subscribers). It could be a terminal to the type of data services pioneered by Ricochet. The cell phone terminal could also bypass the wireless network through wireless local area networks (WLANs). Or, the cell phone terminal could be used as a radio receiver for broadcast programs, a scanner for police frequencies, etc. It is time to think of what we now call the cellphone handset as a future general multi-purpose wireless terminal.

Such multi-purpose terminals would be a threat to most cellular carriers. To see that, let us consider the case of wireless Public and Private Wireless LANs that are emerging on college campuses, airports, office parks, coffee house chains like Starbucks, apartment house complexes, and planes and trains. (cite Bertil Thorngren). These networks, operating on unlicensed spectrum, already reach wireless speeds of up to 11 Mbps two-way communications, and can service, in principle, any type of wireless device, whether laptops, PDAs, pagers, or mobile phones. They follow the 802.111.b standard advanced by Apple, or the Bluetooth standard whose range is more limited.

These new services are a major threat to carriers planning to offer 3G services. In Europe, these carriers are spending vast amounts of money for licenses and infrastructure, yet the most lucrative markets might be cream-skimmed by low-cost, unlicensed, high-speed WLAN providers. It is not too much to state that the very survival of some of the traditional carriers depends on making their 3G investments a success. Their strategies will include offering WLANs themselves. But where that market is open and unconstrained, competition will be a major threat.

The first defensive strategy would be to induce the government to raise the entry barriers by requiring the entire panoply of spectrum licenses, upfront auctions, and national/regional bidding instead of local ones, to raise cost and induce delay.

The second defensive strategy would be to reduce the ability of users to connect into WLANs. This would be done by preventing users to connect their wireless devices into both the carriers' cellular networks and into the WLANs. They could accomplish this by refusing to connect such multi-mode equipment to their networks.

Here, until very recently, the FCC was on their side. For a variety of reasons, the FCC refused to license equipment that could function for multiple different services. As Michael Marcus observes, a ham radio could not be used as a marine radio, etc. As it happens, however, the FCC has reversed this course that had been embarrassingly at variance with its after-lunch rhetoric of flexibility. On September 13, 2001, the FCC adopted rules on software defined radios (SDRs).[3] These rules have moved a great step forward in flexibility. They permit radios whose operating parameters are determined by software—which could be almost any radio these days, if the rules are interpreted generously-- to operate on multiple standards and services.

This initiative has been much to the credit of the FCC. It has the potential to subvert the present carrier-centric regulatory approach. With software-defined radios receiving legal backing, multi-service transceivers will inevitably emerge. Under present rules, however, the carriers would not have to connect such equipment. They ultimately control which equipment can access their networks, and the potential for such equipment to be left stranded will retard its development. This will be developed further below.

   3. *Control Over the Approval of Handsets Reduces Innovation and Choice*

At present, the approval of handsets by carriers and by the FCC is a two- or even three-stage process. The FCC (and similar regulatory bodies elsewhere) issues specifications regarding the radio (RF) and health aspects (SAR)of equipment. This includes frequency, power, and radiation. Some of this is based on self-certification (by way of verification, or declaration of conformity after testing by accredited labs). In other cases, involving higher risk equipment, equipment must be submitted for testing by the FCC or licensed private Telecommunications Certifications Bodies. There are also issues of non-access to certain frequencies for transmission. For example, only licensed pilots can buy transmission equipment for aviation frequencies.

At present, the approval of handsets by carriers and by the FCC is a two- or even three-stage process.

These stages involve the air interface standards that govern the transmission from the handset to the base station, such as CDMA (technical standard IS 95), TDMA (IS 136), I-Den, and GSM. These standards are set by a variety of manufacturer-driven groupings. The decision whether to approve a particular handset for connectivity, however, lies within the discretion of the carrier, since that carrier is entirely free, in the US, to select its standard. In Europe, in contrast, any equipment that complies with the GSM specifications will be connected to the network. There is no carrier discretion. In the US, the industry association CTIA often certifies a manufacturer's equipment to the industry, but each carrier can add its own requirements and flavor of specifications. In consequence, large carriers also test and approve equipment for connection to their network. Hence, the mere adherence by a manufacturer to the standard specifications in the US is not enough. It must also find favor with the carrier. There is no right to use equipment to connect to a cellular network.

---

[3] FCC, Authorization and Use of Software Defined Radios, ET Docket No 00-47.

The carrier's business calculus on what equipment to approve is based on a variety of factors. Since in the US, in contrast to Europe or Japan, the carrier rather than the consumer buy most handsets, low cost is a major factor, as would be serviceability, ability to maintain a limited inventory, and independence from a single source. In addition to reducing the choice available to users, this system also makes manufacturers somewhat dependent on large carriers. They would not lightly put equipment into the marketplace that would be disfavored by the carriers as threatening their basic business by facilitating access to services such as WLAN that compete with the business of their best customers.

### 4. *The Usefulness of User Equipment is Limited by Closed Operating Systems Software*

Beyond the question of whether multi-services equipment can be licensed and connected is the question of control over the nature of the terminals themselves. As handsets become smarter, they begin to resemble small computers. To function, they now begin to incorporate operating system software. At present, there are four initiatives in that direction. Participants include:

- Microsoft (not surprisingly), with its *Stinger*, in cooperation with Samsung and Sendo, an UK-based handset maker.

- Symbian, a joint venture of most cellphone makers resisting a Microsoft dominance, with its *Epoc*. They include Ericsson, Nokia, Motorola, Matsushita, Sony, Psion (software), Kenwood, and IBM.
- *eCos*. This is one of several Linux-based open-source operating system concepts, this one from Red Hat, 3G Labs, and others. At present it is more of a concept than a reality. Manufacturers other than Qualcomm have been reluctant to engage in this non-proprietary environment.
- *J2ME*. (Java version 2, micro edition). This concept, too, encourages openness, though it is not non-proprietary in the way that Linux is.

The important thing to understand is that the choice of the operating system pre-determines the openness of the applications software. With Stinger or Epoc, the carriers and manufacturers would be able to control the types of software applications loaded onto the handsets. With eCos and J2ME, they could not. Hence, open approaches would be preferred by those carriers eager to promote a great diversity of applications. They would favor manufacturers adopting these operating systems. Others, perhaps more concerned—legitimately-- with quality control of applications, security against viruses, and with the increased bargaining strength it gives them over applications developers and third-party applications service providers, would not favor equipment that is open to software, and hence would favor manufacturers who select the closed operating software.

As wireless networks begin to offer increasingly higher-level services, the question of who may load what applications onto a handset, and what network-based service interfaces these applications may access becomes important. Is a user restricted to only the applications that are offered by his primary service provider, or may he load other applications?  Furthermore, can these applications have full access to the functions of the network and the handset?  These questions may be best illustrated with three examples, due to Kevin Kahn.

1. Suppose a brokerage wants to offer a handset application that uses the screen and the alerting (sound or vibration) capabilities of a handset to provide a service to its client.  This application requires that the code at the handset have access to the APIs that access the handset functions and it also requires that the service have access through the network to get messages to that application.  Does deployment of such an application require the cooperation of the wireless service provider?  In today's Internet, deployment of such an application on an end system PC would not require any support from the ISP – will the same be true for wireless?

2. As a second example, suppose a company wishes to deploy a universal messaging application that can alert users to any messages, email or voice, that they have on any of a number of message queuing services.  This application requires access to the alerting indicators of the handset and it may also require access to the voicemail service inherent in the wireless providers service so that messages queued there can be included in the new service.  Do the necessary APIs and addressing paths exist to allow such an application.

3. As a final example, suppose a vending machine with an infrared or Bluetooth interface that wants to interact with a handset to communicate with a backend billing service to handle the vending.  This requires a digital certificate to be sent from the machine through the handset to the backend service followed by an authorization certificate being sent back via the same path to the machine.  Again, this requires the application to have access to specific functions of the handset (probably the IR port, the screen to present selections, and the buttons to select). Can this application be deployed without cooperation of the wireless service provider?

## 5. *Closed Portals Reduce User Choice*

Under the presently evolving system, the user reaches a wireless portal, whence she can be connected to a variety of other sites. The selection and placement of these links, however, is under the control of the carrier. Other portals might be accessed, but that requires additional clicks. This situation is very similar to the one discussed for cable television's access to portals other than those of the cable company or its partner. It has led, in the US, to requirements on Time Warner Cable to open its system to several service providers/portals. Similar rules are being contemplated for the entire cable industry. These issues are well known and require no recapitulation. Virtually the same arguments on both sides apply also to cell phone access to portals, and through them, to the broader internet. It should be noted, however, anticipating the conclusion of this paper, that they are much easier to resolve to the wireless medium.

## 6. *Transactions and Content are Limited.*

The wireless carrier's portal is not a common carrier. Hence, the selection of websites, e-vendors, and content providers is entirely that of the carrier. Its selection would be based on its own economic, cultural, and political considerations. Being a selector, it would also incur some legal liability, which would further increase caution.

Even where there might be a wide set of links, such as in the case of Japan's NTT DoCoMo and its i-mode portal, such openness can be affected by differentiating policies. DoCoMo has preferential arrangements with a small number of its partner sites, and handles their billing through the users' phone bill. In contrast, the majority of other sites must arrange for their own billing, putting them at some disadvantage. This provides DoCoMo with some leverage over its primary partners.

The absence of openness resembles the "walled garden" arrangements of some Internet portals provided by cable companies. Correspondingly, we can term this arrangement the "walled airwave" system.

**Implications for Public Policy**

The previous section has identified the potential for real problems. But the recognition of such issues does not mean that regulatory approaches are needed. A vigorous competition among mobile carriers could overcome most issues and generate unbundling through market forces. At the same time, the ability to exercise market power with respect to mobile commerce providers or wireless LANs might be common to all mobile providers and more profitable than a more open system. In such a case, market forces might not lead to unbundling.

The knee-jerk response to the problems identified in this paper is that competition will take care of it. But suppose that carriers would be consistently worse off by offering consumers the choice of moving easily around to other carriers or service providers. Such competition would reduce prices and profitability. It would, on the other hand, grow the market. But it is quite likely that each carrier would be better off servicing a less competitive slice of a smaller market, rather than engaging in greater competition in a larger market.

It is not clear why a carrier A would be the first to offer such choice to its customers. After all, it would provide an exit to its own customers, without a potential compensating gain from the customers of the other carriers B and C. The main reason would be to hope for enough users of B and C to switch their subscriptions to A in order to have the choice of *not* using A. This can hardly be a strong selling point. Furthermore, any choice requires the consent and cooperation of B and C, which might not be forthcoming once they realize that they are opening the door to a mutually destabilizing competition. They will be concerned with reputation effects if they are blamed in users' mind with poor performance caused by an element not under their direct control. And they might be able to use bundling as a way to price discriminate, as George Stigler has pointed out in a different context. The likelihood of oligopolistic behavior within a small group of carriers is high. As the number of competitors shrinks, each has less to gain and more to lose by maverick behavior. It is also an inhibitor for any software developer to take initiatives for new applications if the market is largely closed, and this further reduces the attractiveness of any non-conforming behavior be a carrier.

Where market forces do not work, would regulation? Let us look at several potential points of intervention and evaluate their need.

A schematic view of an unbundled wireless network environment is provided in Graph 2. It shows, at each stage of the chain of wireless provision, alternative providers. We will consider "openness" at several points, and conclude that only one of them – the openness of the terminal equipment to access multiple providers of wireless services and providers-- is critical. A subsidiary second opening – unlicensed spectrum—supports such policy. This is discussed in the following.

**Graph 2**

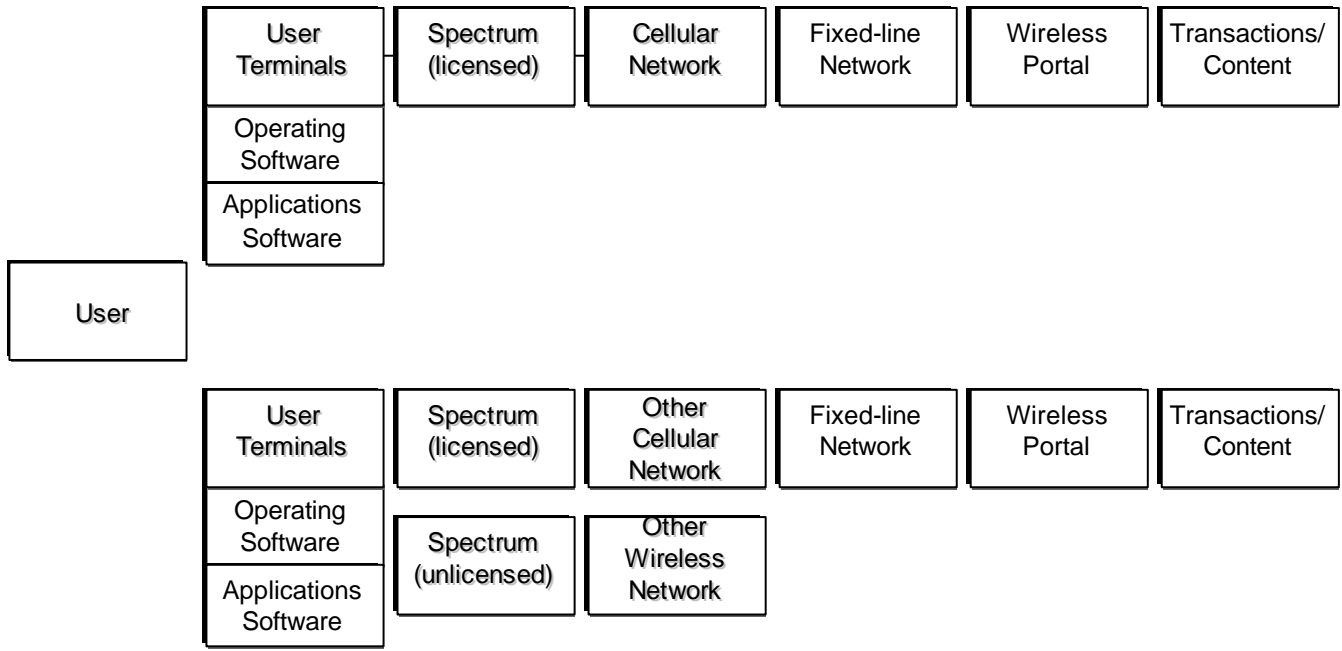

## 1. *The Separation of the User Equipment (UE) From the Carrier*.

Such a policy would amount to a '*Carterfone*' policy for users' wireless equipment. While the carrier could still offer and market its preferred equipment, it could not exclude other equipment, as long as it conforms to certain technical specifications pertaining to the RF transceiving function and non-discriminatory industry specifications for air interfaces standards. These specifications could not close equipment third-party applications or access to other network protocols offered by other types of providers, as long as it conforms to the FCC's software defined radio rules.

While a fully bundled service could be offered by a carrier as before, the carrier could not prevent a user of selecting, for any given call, another wireless service provider or use the equipment for other communications purposes.

The significance of such arrangement is that equipment will be offered by the market that adds features, and, more importantly, permits a user to select service providers depending on circumstances. For example, a user in a shopping mall, campus, office building, or airport could connect to a wireless LAN. A user encountering a circuit busy could switch to another carrier. A user seeking to receive synchronous music, radio style, could do so by accessing a specialized broadcaster.

This choice would reduce the need for most other access requirements, since the user would not be tied to a single carrier with significant costs of switching to another. This is partly embodied in the GSM standard which provides some user selectivity over carriers, though the approval of such alternatives remains with the primarily carrier, which also handles the billing.

This approach would be similar to that adopted by the FCC for CPE following the *Carterfone* decision in 1968. It followed Cassandra warnings of impending network chaos, but has worked spectacularly well.

   2. ***Access to Unlicensed Spectrum***.

The key source of leverage for carriers is the high entry barrier for new and future entrants in service provision, due to the spectrum auctioning system with its advance payment feature. Given the difficulty in freeing additional spectrum and the high cost of aquiring it, it seems unlikely that there would be new entrants emerging to challenge the reduced group of carriers. Therefore, government should additionally provide adequate spectrum on a license-free basis, with users and service providers paying for usage rather than for ownership, in the way that automobiles pay for the use of highways. This has been developed in detail by the author in other papers.[4] Once such spectrum is available, and once users' terminals can access service providers such as WLANs operating on such spectrum, users will not be constrained by the limited choice of maybe four cellular carriers that could still collectively be restrictive.

### 3. Access to Alternative Wireless Portals.

The third access issue is that to the wireless portal. The issues here are similar to those discussed for the cable industry. The similar arrangement would mean that the wireless carrier would let the user pre-select its primary portal, or that several such portals would be accessible at no extra effort, or that the two upper layers of the carrier portal would be open to third parties. This approach would mirror the open access of the internet, and the approaches now being applied to AOL Time Warner, and considered by the FCC in its proceedings.

Content openness may be the easiest type of openness to consider since it is essentially a browser level openness. The question can be reduced to whether the user can enter an arbitrary URL to a network portal to access content (independent of any business deal between the wireless provider and particular content providers) and whether browser plug-ins can be created and downloaded to render

---

[4] See Noam, Eli M., "Spectrum Auctions: Yesterday's Heresy, Today's Orthodoxy, Tomorrow's Anachronism. Taking the Next Step to Open Spectrum Access," *The Journal of Law & Economics*, vol. XLI part 2 pp. 765-790

the resulting content if required. This issue is analogous to the walled garden discussions that have occurred in the wired internet.

This problem would largely go away if the users could access, through their handsets linked to other carriers and wireless providers, also other portals and websites.

### 4. Openness of the carrier's network

The fourth element of openness relates to services offered by third parties and requiring presence in the wireless network itself. The options are either to keep wireless networks closed to third parties, or total openness, resembling a common carrier access for third party software applications, or a type of equipment collocation that exists in telecommunications. Here, too, would the ability to access alternative wireless carriers through flexible handsets be enough to deal with this issue.

## Conclusion

The focus of the FCC policy has been to provide carriers with choice—in the utilization of he licensed frequency, in the technical specifications of it service, in its pricing, etc. There does not seem to have been a similar orientation towards choice of the users, broadly defined as consumers and providers of various attached services. The implicit notion was that by providing carriers with options, and creating competition, users will be well served. And that certainly goes a long way. But carriers are likely to resist offering consumers the choice of moving

---

(October 1998).

easily around to other carriers or service providers. Such competition would reduce prices and profitability.

The conclusion of the analysis is that the key point of openness, and arguably the only one needed, is that of *openness of user equipment.* With this openness achieved, the user would have alternative avenues to spectrum, content, portals, etc. A secondary policy would be to assure alternative wireless pathways such as WLANs by providing an adequate amount of unlicensed spectrum.

Why is all this important? The overall goal of the openness approach described above is to establish for the wireless environment the same dynamism shown in the internet, encouraging hardware and software innovation and applications. Right now this is a dynamic sector, mostly based on the growth of penetration. Soon, however, this growth will plateau as universal wireless connectivity is being approached. At that point, we need the impetus for further innovation that a more open system provides. For the carriers, the overall positive impact in terms of traffic generation may well outweigh some loss of control. For users, service providers, and technology developers, the advantages of openness might be significant.

American communications policy has fared best when it puts its faith in the dynamism of the periphery of the network, instead of seeking to strengthen the ability of the network core to dominate. Wireless is no exception. And the mediocre results of policies focusing on the core, in contrast to those for other parts of the communications environment, suggest that a reorientation is in order. The key step now is to follow the opening set by the FCC's for software defined radio by a *Carterfone*-style opening to equipment that can access multiple wireless networks.

**Acknowledgements**

I am grateful for the help and comments received by:
James Alleman, Bob Atkinson, Ron Barnes, Brian Bebchick, Kenneth R. Carter, Kathryn Condello, Terry Hsiao, John Lee, Don Nichols, Michael Noll, Michael Marcus, Bertil Thorngren, John Williams, and to Charlie Firestone and the Aspen Institute's Regulatory Policy Meeting, especially Kevin Kahn and Robert Pepper. Views expressed here are entirely my own.